\documentclass[conference]{IEEEtran}

\usepackage{epsfig}

\newcommand{\CK}{\v Cerenkov}

\begin{document}

\title{Particle identification with the AMS-02 RICH detector: $D$/$p$ and $\bar{D}$/$\bar{p}$ separation}

\author{\authorblockN{Lu\'isa Arruda, Fernando Bar\~ao,
\underline{Rui Pereira}}
\authorblockA{LIP/IST \\
         Av. Elias Garcia, 14, 1$^{\textnormal{\scriptsize{o}}}$ andar\\
         1000-149 Lisboa, Portugal \\
         e-mail: pereira@lip.pt}}

\maketitle

\begin{abstract}

The Alpha Magnetic Spectrometer (AMS), whose final version AMS-02 is to be
installed on the International Space Station (ISS) for at least 3 years,
is a detector designed to measure charged cosmic ray spectra with energies
up to the TeV region and with high energy photon detection capability up
to a few hundred GeV, using state-of-the art particle identification
techniques.

Among several detector subsystems, AMS includes a proximity focusing RICH
enabling precise measurements of particle electric charge and velocity.
The combination of both these measurements together with the particle
rigidity measured on the silicon tracker endows a reliable measurement of
the particle mass.

The main topics of the AMS-02 physics program include detailed
measurements of the nuclear component of the cosmic-ray spectrum and the
search for indirect signatures of dark matter. Mass separation of singly
charged particles, and in particular the separation of deuterons and
antideuterons from massive backgrounds of protons and antiprotons
respectively, is essential in this context. Detailed Monte Carlo
simulations of AMS-02 have been used to evaluate the detector's
performance for mass separation at different energies.  The obtained
results and physics prospects are presented.

\end{abstract}

\section{The AMS-02 experiment}

\begin{figure}[htb]

\center

\vspace{0.2cm}

\mbox{\epsfig{file=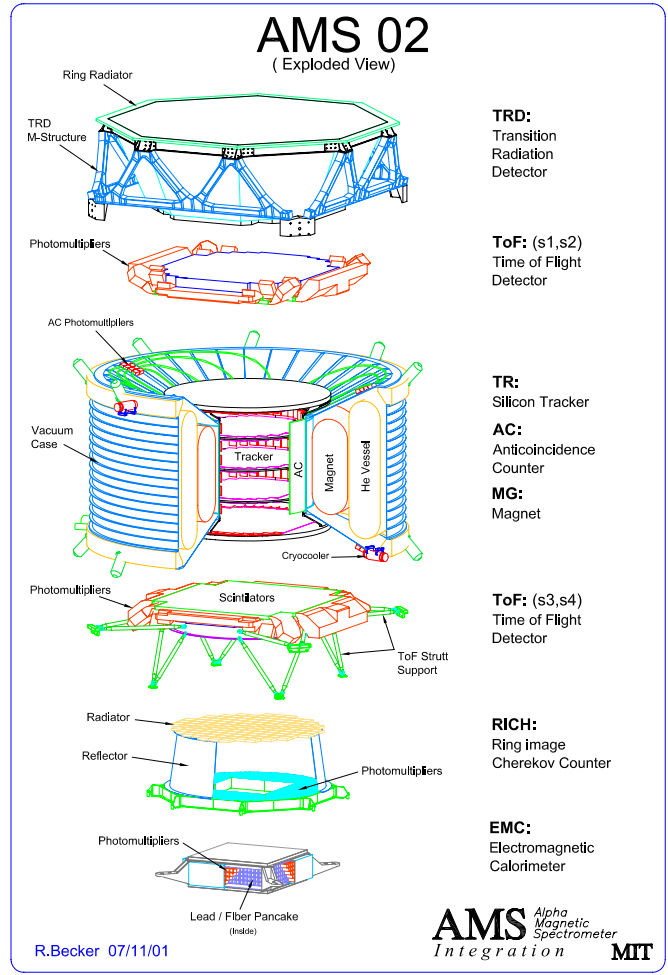,width=0.48\textwidth,clip=}}


\caption{Exploded view of the AMS-02 detector.\label{amsdet}}

\vspace{-0.5cm}

\end{figure}

The Alpha Magnetic Spectrometer (AMS)\cite{bib:ams}, whose final version
AMS-02 is to be installed on the International Space Station (ISS) for at
least 3 years, is a detector designed to study the cosmic ray flux by
direct detection of particles above the Earth's atmosphere using
state-of-the-art particle identification techniques. AMS-02 is equipped
with a superconducting magnet cooled by superfluid helium. The
spectrometer is composed of several subdetectors: a Transition Radiation
Detector (TRD), a Time-of-Flight (TOF) detector, a Silicon Tracker,
Anticoincidence Counters (ACC), a Ring Imaging \CK\ (RICH) detector and an
Electromagnetic Calorimeter (ECAL). Fig.~\ref{amsdet} shows a schematic
view of the full AMS-02 detector. A preliminary version of the detector,
AMS-01, was successfully flown aboard the US space shuttle Discovery in
June 1998\cite{bib:ams01res}.

The main goals of the AMS-02 experiment are:

\begin{itemize}

\item A precise measurement of charged cosmic ray spectra in the rigidity
region between \mbox{$\sim$ 0.5 GV} and \mbox{$\sim$ 2 TV}, and the detection
of photons with energies up to a few hundred GeV;

\item A search for heavy antinuclei ($Z \ge$ 2), which if discovered would
signal the existence of cosmological antimatter;

\item A search for dark matter constituents by examining possible
signatures of their presence in the cosmic ray spectrum.

\end{itemize}

The long exposure time and large acceptance (0.5 m${}^2\cdot$sr) of
AMS-02 will enable it to collect an unprecedented statistics of more than
$10^{10}$ nuclei.

\section{The AMS RICH detector}

One of the subdetectors in AMS-02 is a proximity focusing Ring Imaging
\CK\ (RICH) detector. It is composed of a dual radiator with silica
aerogel ($n=$~1.050) and sodium fluoride ($n=$~1.334), a high reflectivity
lateral conical mirror and a detection matrix with 680 photomultipliers
coupled to light guides.

\begin{figure}[htb]

\center


\mbox{\epsfig{file=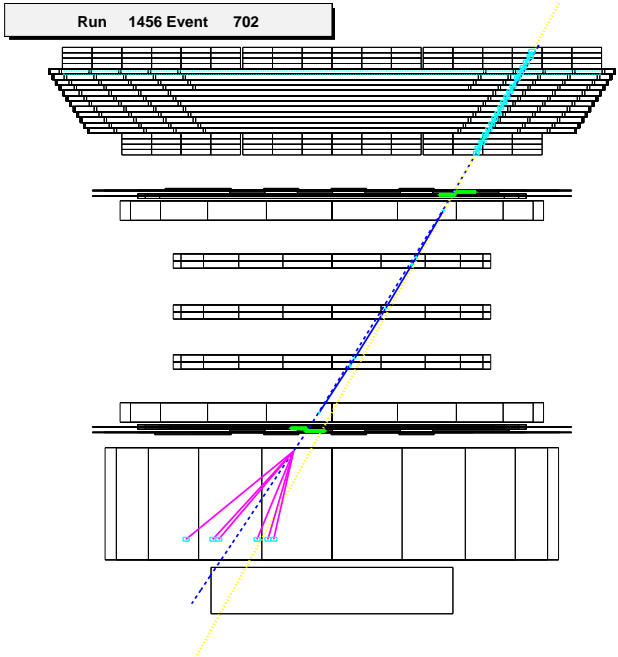,width=0.48\textwidth,clip=}}

\vspace{-0.3cm}

\caption{A simulated proton event as seen in the AMS-02 display.\label{amsdisplay1}}

\end{figure}

\begin{figure}[htb]

\center

\vspace{-0.5cm}

\mbox{\epsfig{file=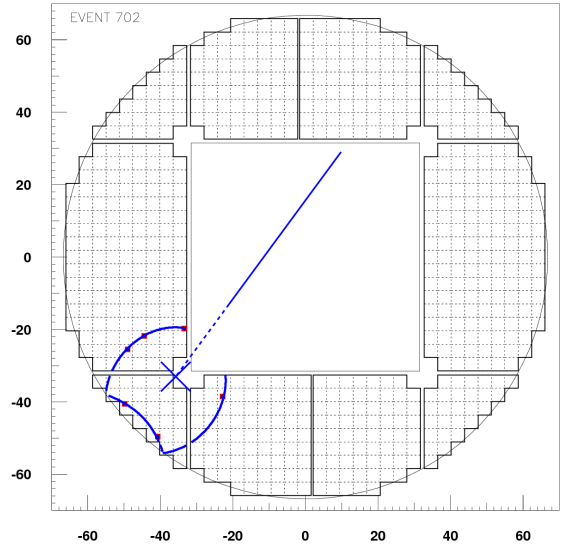,width=0.48\textwidth,clip=}}

\vspace{-0.3cm}

\caption{The same event of Fig.~\ref{amsdisplay1} as seen in the
RICH display developed at LIP.\label{amsdisplay2}}

\end{figure}

The RICH detector will provide a very accurate velocity measurement (in
aerogel, $\Delta \beta / \beta \sim$ 10${}^{-3}$ and 10${}^{-4}$ for
$Z=$~1 and $Z=$~10~$-$~20, respectively) and charge identification of
nuclei up to iron ($Z=$~26).

RICH data, combined with information on particle rigidity from the AMS
Silicon Tracker, enable the reconstruction of particle mass. A typical
RICH event is shown in Figs.~\ref{amsdisplay1} and \ref{amsdisplay2},
where the latter gives a detailed view of the readout matrix. The accuracy
of the RICH velocity measurement is essential due to the growth of
relative errors when $v \to c$:

\begin{displaymath}
\frac{\Delta m}{m} = \frac{\Delta p}{p} \oplus \gamma^2 \frac{\Delta
\beta}{\beta}
\end{displaymath}

The assembly of the AMS RICH detector is currently underway at CIEMAT in
Madrid. The integration of the RICH and the other subdetectors of AMS-02
will take place at CERN in 2007.

The analysis of RICH data involves the identification of the \CK\ ring in
a hit pattern which usually includes several scattered noise hits and an
eventual strong spot in the region where
the charged particle crosses the detection plane. Two independent algorithms
for velocity and charge reconstruction have been developed in the AMS
collaboration for the analysis of RICH events: a geometrical method based
on a hit-by-hit reconstruction\cite{bib:elisa}, and a method using all the
hits with the maximization of a likelihood function\cite{bib:rich2003}).

A prototype of the RICH detector, consisting of 96 photomultiplier units,
was tested both with cosmic ray particles and with beam ions at the CERN SPS
in 2002 and 2003. A piece of the conical reflector was included in the
beam test setup\cite{bib:prototype}. The algorithms for velocity and charge
reconstruction were successfully applied to data from these prototype
tests\cite{bib:phdluisa}.

\section{Components of the cosmic ray spectrum}

Protons are the most abundant component ($\sim$~90\%) of charged cosmic rays
reaching the Earth's vicinity. The remaining fraction is essentially made of
atomic nuclei (mostly ${}^4$He) and single-charged particles such as $e^-$,
$e^+$ and $\bar{p}$.

Precise measurements of the smaller components in the cosmic ray spectrum
are essential in the context of the study of cosmic-ray production and
acceleration. The required precision can only be attained through very
effective charge and mass discrimination methods since the abundances of
different components differ by several orders of magnitude.

Ratios such as D/p, ${}^3$He/${}^4$He and B/C give information on the
interstellar medium since all compare the abundances of secondary and
primary species. The beryllium isotope ratio ${}^{10}$Be/${}^9$Be is a probe
for galactic confinement times since both isotopes are secondary but one
of them, ${}^{10}$Be, is unstable ($t_{1/2}=$~1.5 $\times$ 10$^6$ yr).

\begin{figure}[htb]

\center


\mbox{\epsfig{file=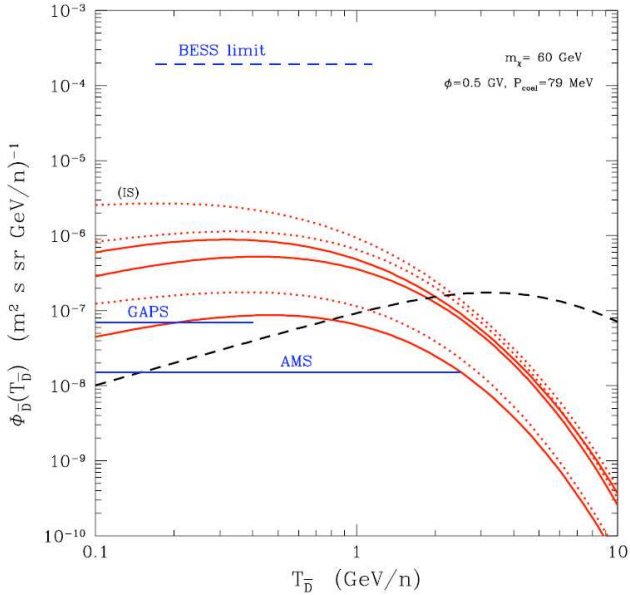,width=0.48\textwidth,clip=}}

\vspace{-0.3cm}

\caption{
Comparison between expected antideuteron flux from secondary processes
(dashed line) and the flux from the annihilation of a 60 GeV dark matter
particle (solid lines: solar minimum; dotted lines: interstellar flux).
Fluxes from dark matter annihilation are shown for three sets
of propagation parameters.
(from Ref.~\protect \cite{bib:donformau07}) \label{antidsusy}}


\end{figure}

The lightest neutralino ($\tilde{\chi}^0_1$), predicted by supersymmetric
models, is a strong dark matter candidate. If it exists and accounts for the
known dark matter density ($\Omega_{CDM} \simeq \Omega_m - \Omega_b \simeq 0.2$)
or at least for a significant part of it, neutralino annihilation
($\tilde{\chi}^0_1 \tilde{\chi}^0_1 \to ...$) must take place and
contribute to the observed cosmic ray composition, with the more visible
effects occurring in the spectra of antiparticles like $e^+$, $\bar{p}$ and
especially $\bar{D}$\cite{bib:donformau07}. Fig.~\ref{antidsusy} shows
a comparison between the expected $\bar{D}$ fluxes from secondary
production and from dark matter annihilation.

\section{Particle identification}

To evaluate the capabilities of AMS-02 for mass separation of deuterons and
antideuterons from other particles with the same charge, studies have been
performed using the case of deuteron vs. proton separation. In the
past, studies on the separation of helium ($Z=$~2) and beryllium ($Z=$~4)
isotopes have also been performed using a standalone simulation of the
RICH detector\cite{bib:mscluisa}. In the present case the large difference
between proton and deuteron abundances (D/p $\sim$ 1\%) increases
the importance of a very effective mass separation to isolate the deuteron
signal from a large background of proton events.

In the study of D/p separation a full-scale simulation of the AMS detector
was used. Particles were simulated as coming from the top plane of a cube,
corresponding to an acceptance of 47.78 m${}^2\cdot$sr. Three data samples were
chosen. Table \ref{simstat} shows the momentum ranges and number of events
simulated in each sample.

\begin{table}[htb]


\begin{center}

\caption{Samples used in the $D/p$ separation studies}
{
\begin{tabular}{|c|c|c|}

\hline \textbf{Sample} & \textbf{Momentum range} & \textbf{No. events}
\\ \hline $p$ (low momentum) & 0.5~$-$~10 GeV/c & 3.1~$\times$~10${}^8$
\\ \hline $p$ (high momentum) & 10~$-$~200 GeV/c & 1.3~$\times$~10${}^8$
\\ \hline $D$ & 0.5~$-$~20 GeV/c & 5.6~$\times$~10${}^7$

\\ \hline \multicolumn{3}{c}{}

\end{tabular}

\label{simstat}
}

\end{center}


\end{table}

\begin{figure}[htb]

\center


\mbox{\epsfig{file=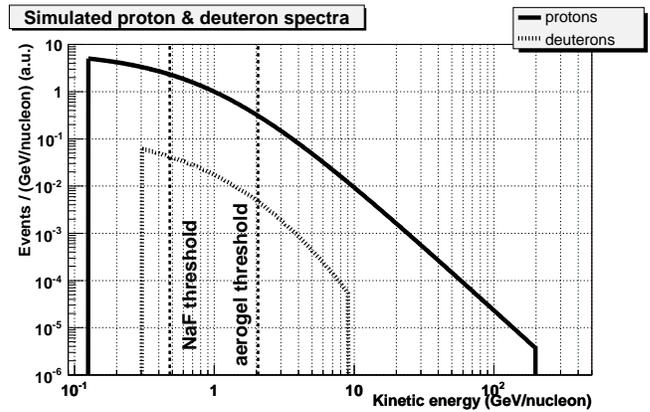,width=0.48\textwidth,clip=}}

\vspace{-0.3cm}

\caption{Simulated proton and deuteron spectra used in this work.
\label{simspec}}


\end{figure}

For each sample, $\frac{dN}{d(ln \, p)} =$~constant. Variable
weights were assigned to events in order to compensate for the statistics
in each sample and to reproduce a realistic spectrum (Fig.~\ref{simspec}):

\begin{itemize}

\item The simulated proton spectrum followed \mbox{$dN/dE \propto E^{-2.7}$};

\item The simulated deuteron spectrum was calculated combining the proton
spectrum above with D/p ratios taken from Ref.~\cite{bib:seo:h}.

\end{itemize}

In each event a set of preliminary data selection cuts using readings from
different subdetectors of AMS-02 was applied to reduce the fraction of
events with a bad reconstruction. Only downgoing events ($\beta > 0$)
were accepted. In addition, events were accepted if the following conditions
were satisfied:

\begin{itemize}

\item Only one particle was detected in the event;

\item A particle track was reconstructed by the Silicon Tracker;

\item No clusters were found in the Anti-Coincidence Counters;

\item Clusters from at least 3 TOF planes (out of 4) were used for event
reconstruction;

\item At most one additional cluster was allowed in the TOF; 

\item At least 6 Tracker layers (out of 8) were used in the track
reconstruction;

\item Compatibility was required for the rigidity measurements obtained from
two different algorithms, with \mbox{$\Delta R/R <$~3\%};

\item Compatibility was also required for the rigidity measurements obtained
from each half of the Tracker (upper and lower), with
\mbox{$\Delta R/R <$~50\%};

\item The particle's impact point on the RICH radiator was less than 58 cm
from the centre (i. e. more than 2 cm from the mirror);

\item At most one track was present in the TRD;

\item The TOF and Tracker charge reconstructions were compatible.

\end{itemize}

Among the events that triggered the detector, a fraction corresponding to
\mbox{$\sim$~15-20\%} of proton events and \mbox{$\sim$~10-15\%}
of deuteron events in the relevant region of kinetic energy (few GeV/nucleon)
passed this set of preliminary cuts, corresponding to an acceptance of
$\sim$~0.3 m${}^2\cdot$sr for protons and $\sim$~0.2 m${}^2\cdot$sr for
deuterons.

The reconstruction of particle masses was then performed for events having
a signal in the RICH detector. The extremely accurate velocity measurement
provided by the RICH ($\Delta \beta / \beta \sim$ 10${}^{-3}$ in the case
of protons and deuterons) is crucial to reduce the background level. A series
of event selection cuts were introduced, based on data provided by the RICH
and the results of the two reconstruction algorithms:

\begin{figure}[htb]

\center


\mbox{\epsfig{file=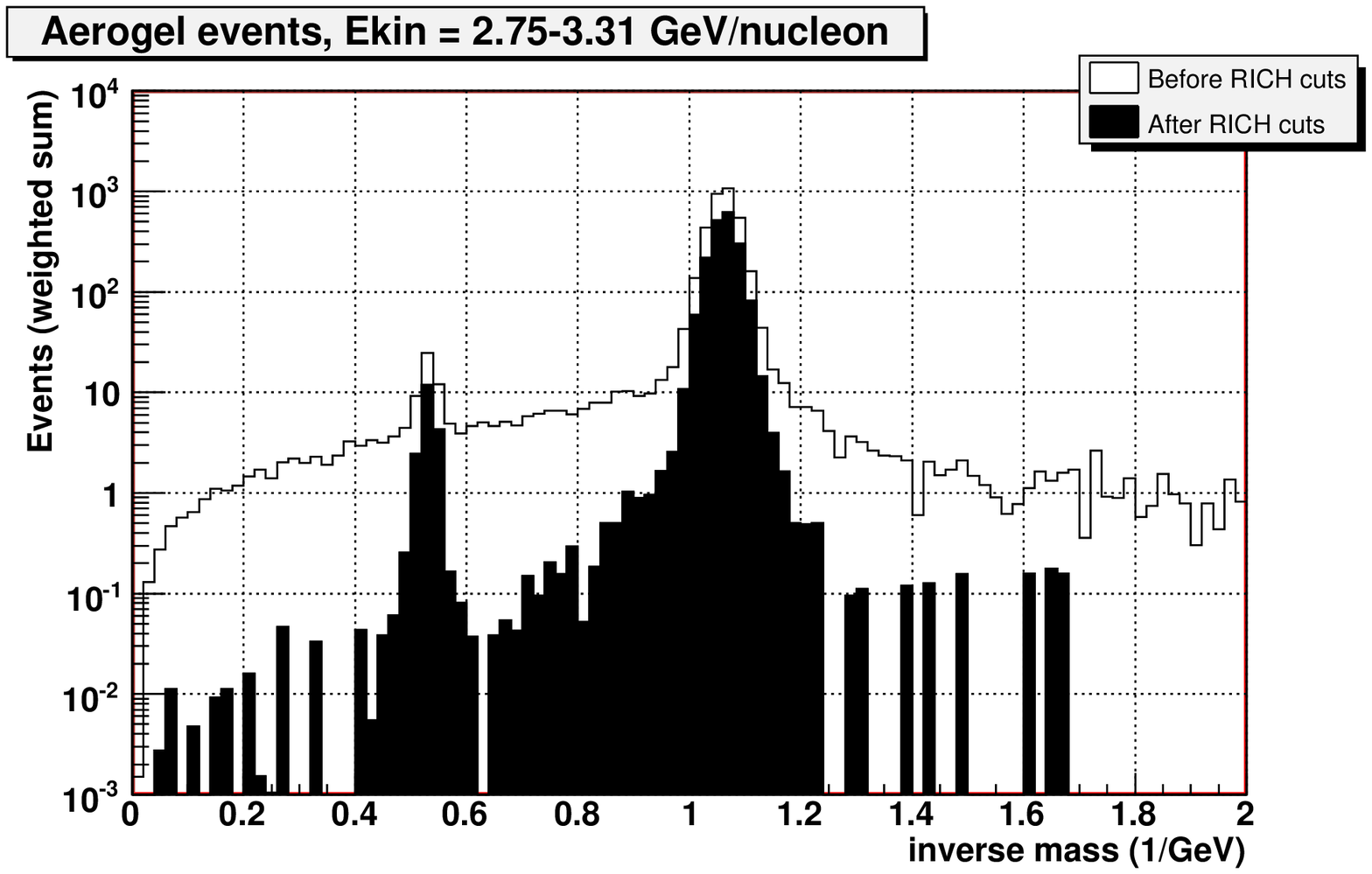,width=0.48\textwidth,clip=}}

\mbox{\epsfig{file=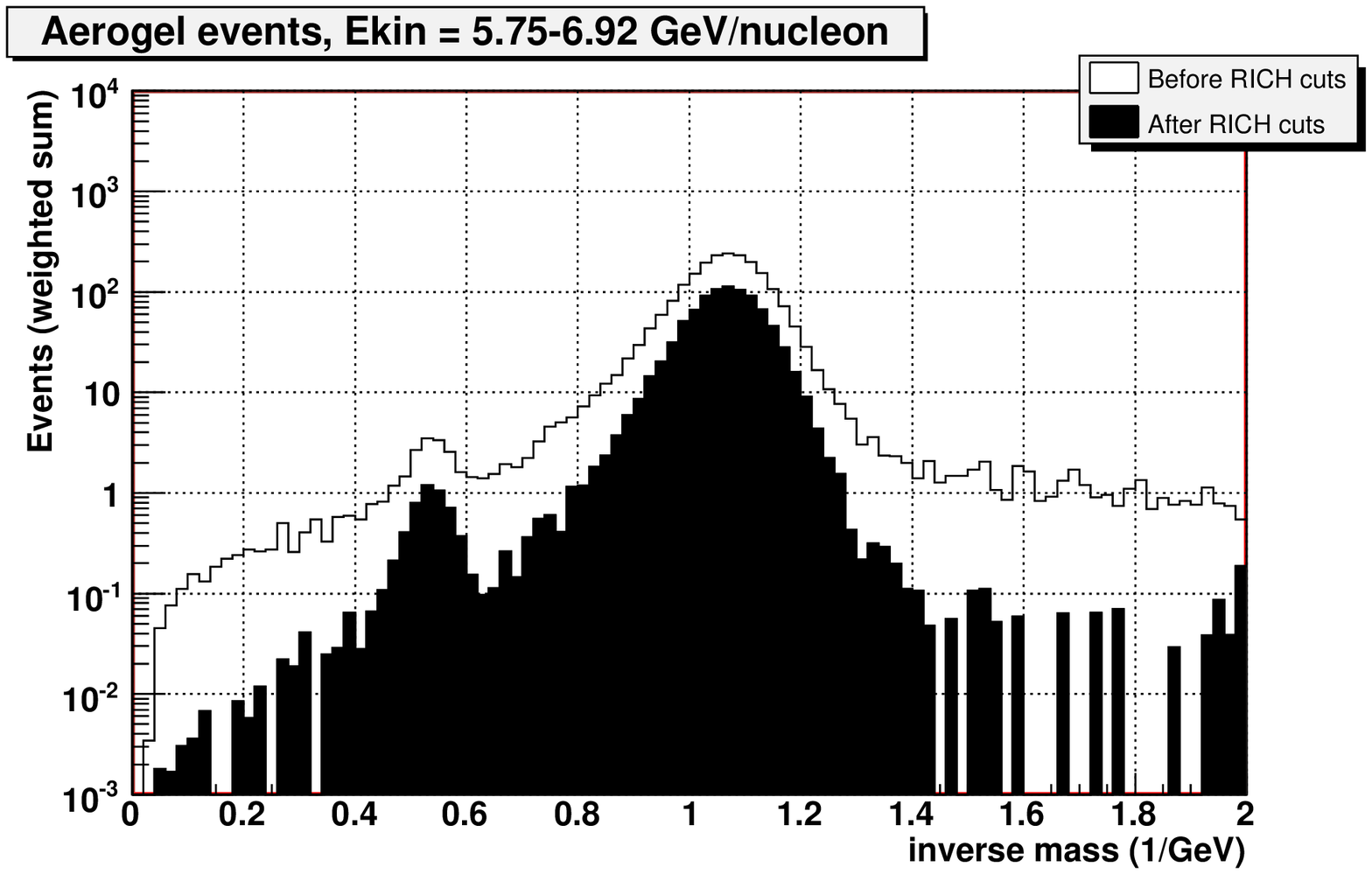,width=0.48\textwidth,clip=}}

\vspace{-0.3cm}

\caption{Examples of inverse mass distribution in aerogel events for two
energy regions.
\label{imassagl}}

\end{figure}

\begin{figure}[htb]

\center


\mbox{\epsfig{file=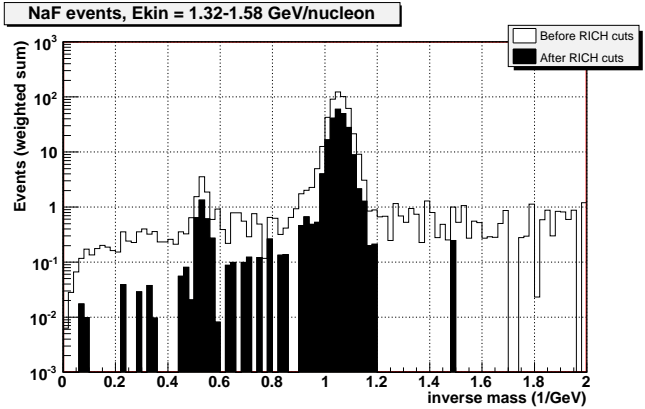,width=0.48\textwidth,clip=}}

\vspace{-0.3cm}

\caption{Example of inverse mass distribution in NaF events.
\label{imassnaf}}

\end{figure}

\begin{figure}[htb]

\center


\mbox{\epsfig{file=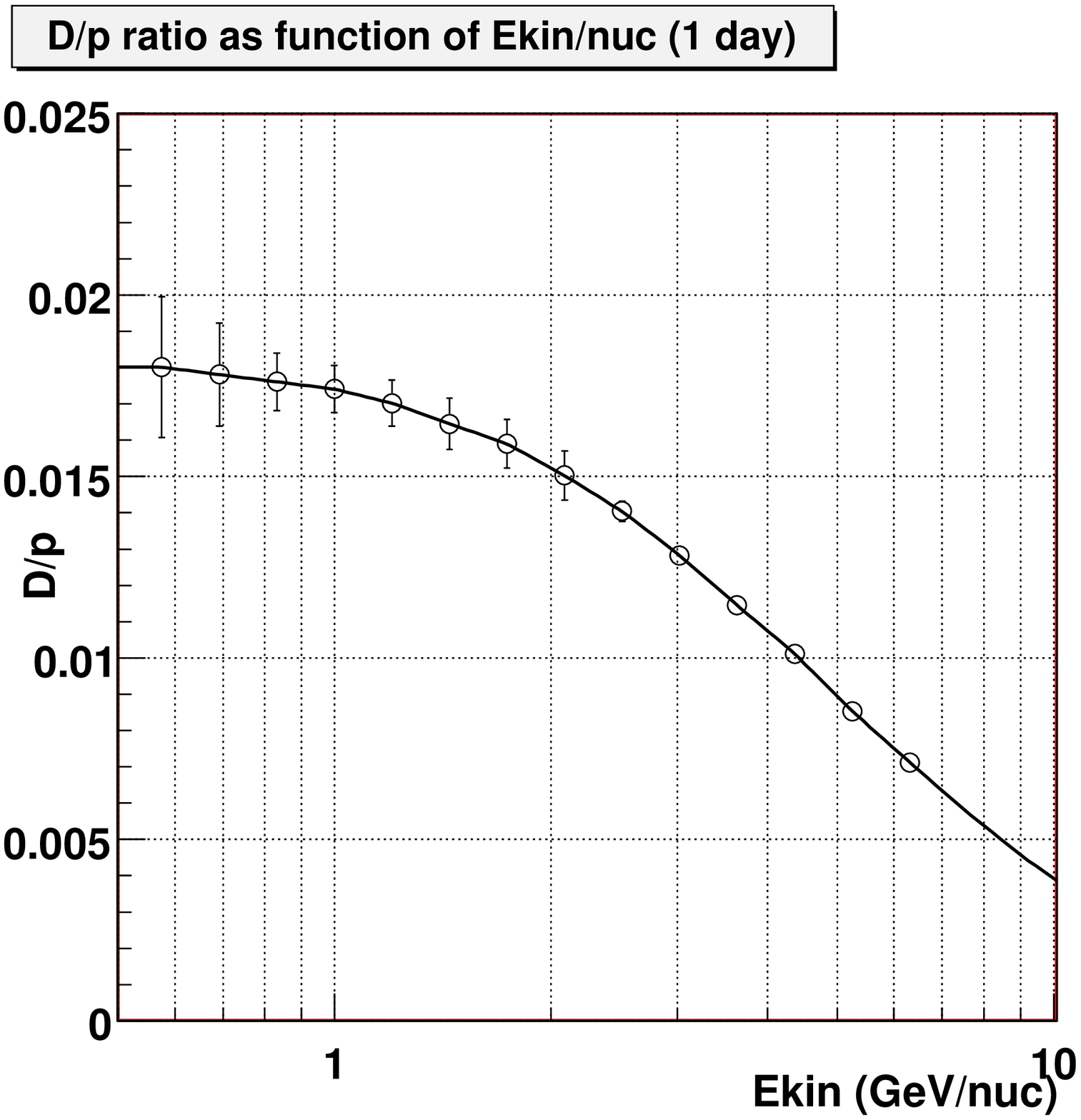,width=0.48\textwidth,clip=}}

\vspace{-0.3cm}

\caption{Expected sensitivity of AMS for D/p ratio with one day of data.
\label{dpratio1day}}


\end{figure}

\begin{itemize}

\item A \CK\ ring was reconstructed using each method, and at least 3 hits
were used in both cases;

\item The total ring signal was not higher than 10 photoelectrons in NaF
events, and not higher than 15 photoelectrons in aerogel events;

\item A Kolmogorov test to the uniformity of the hits azimuthal distribution
in the ring gave a result of at least 0.2 in the case of NaF events, and 0.03
in the case of aerogel events;

\item Compatibility was required for the velocity measurements from the TOF
and RICH detectors, with \mbox{$\Delta \beta/\beta < 10\%$};

\item Compatibility was also required for the velocity measurements obtained
from the two RICH reconstruction methods, with \mbox{$\Delta \beta/\beta < 0.3\%$}
for NaF events, and \mbox{$\Delta \beta/\beta < 0.1\%$} for aerogel events;

\item The reconstructed, rounded electric charge obtained from the geometrical
method was 1 or 2;

\item The reconstructed, non-rounded electric charge obtained from the
likelihood method was between 0.5 and 1.5 in NaF events, and between 0.6
and 1.4 in aerogel events;

\item The ring acceptance (visible fraction), as estimated by
the likelihood method, was at least 20\% in NaF events, and at least 40\% in
aerogel events;

\item The number of noisy hits not associated to the crossing of the charged
particle (i. e., hits that were far from the reconstructed ring and far from
the estimated crossing point of the charged particle in the detection matrix)
was not higher than 2 in NaF events, and not higher than 3 in aerogel events.

\end{itemize}

\section{Analysis results}

Results show that mass separation of particles with $Z=$~1 is feasible even
if one species is orders of magnitude more abundant than the other. D/p
separation is possible up to $E_{kin} \sim$~8 GeV/nucleon. Some examples
of the mass distributions obtained are shown in Figs.~\ref{imassagl} and
\ref{imassnaf}. Solid lines show the mass distributions before the RICH
cuts were taken into consideration.

Fig.~\ref{dpratio1day} shows the expected sensitivity of AMS for the D/p
ratio after one day of data taking. Results show that a single day of AMS-02
statistics will be sufficient to improve on the existing data for this ratio.

\begin{figure}[htb]

\center


\mbox{\epsfig{file=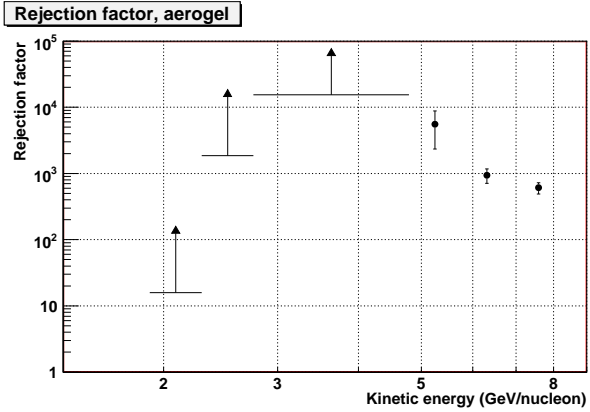,width=0.48\textwidth,clip=}}

\vspace{-0.3cm}

\caption{Rejection factor for D/p separation in aerogel events.
\label{nrejfac}}

\end{figure}

\begin{figure}[htb]

\center


\mbox{\epsfig{file=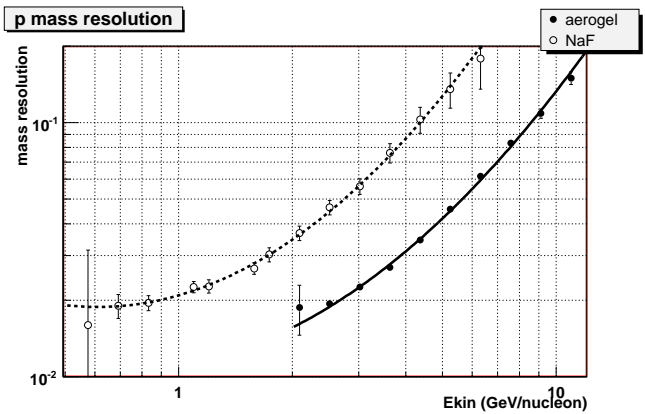,width=0.48\textwidth,clip=}}

\vspace{-0.3cm}

\caption{Relative mass resolution for protons: NaF events (open dots)
and aerogel events (filled dots).
\label{massres}}


\end{figure}

In the optimal region immediately above the aerogel
radiation threshold ($E_{kin} <$~5 GeV/nucleon) rejection factors
higher than 10${}^4$ region were attained (Fig.~\ref{nrejfac}). The
best relative mass resolutions for protons (Fig.~\ref{massres}) and deuterons
are $\sim$~2\% for both radiators in the regions above their respective
thresholds.

\begin{figure}[htb]

\center


\mbox{\epsfig{file=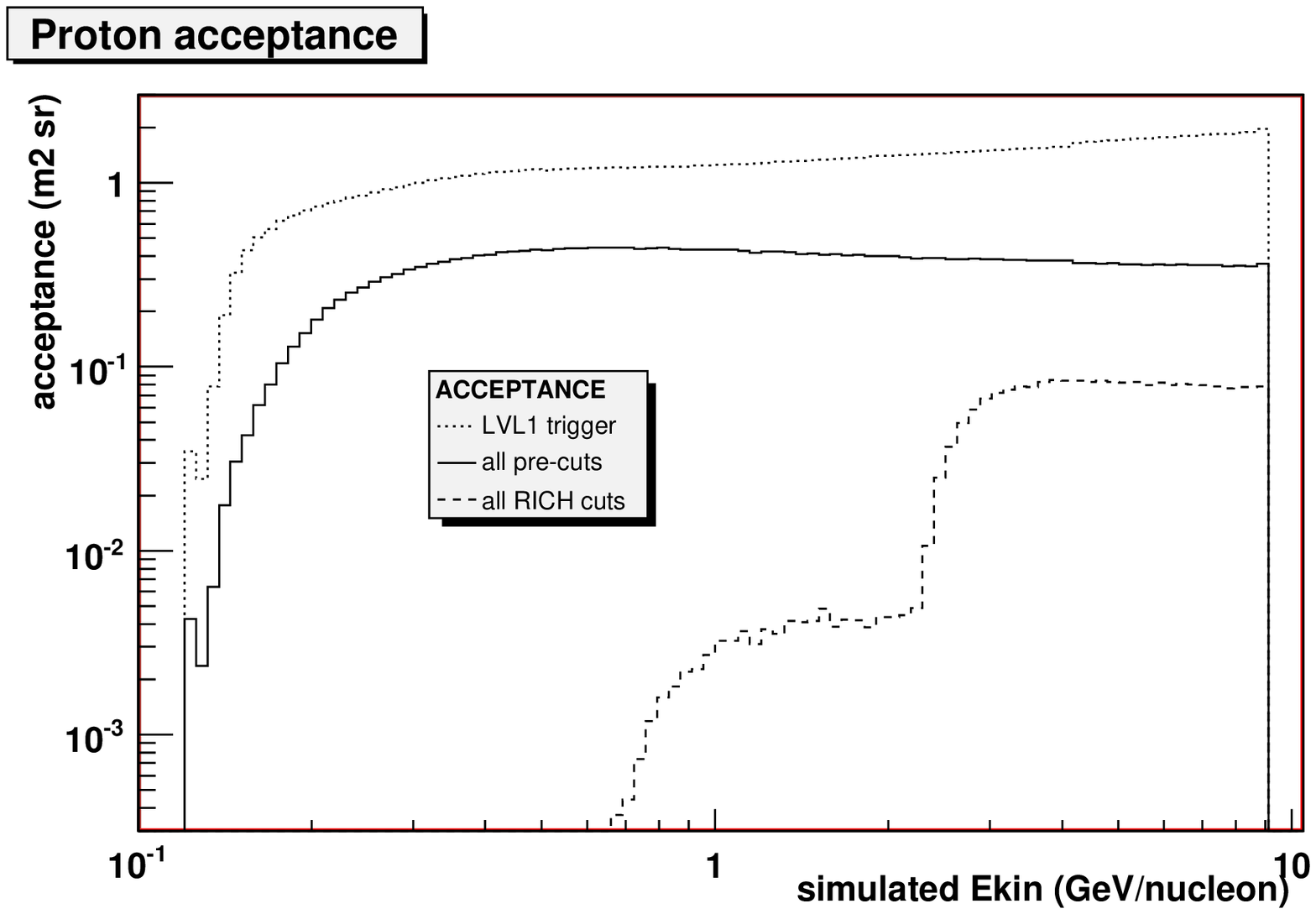,width=0.48\textwidth,clip=}}

\mbox{\epsfig{file=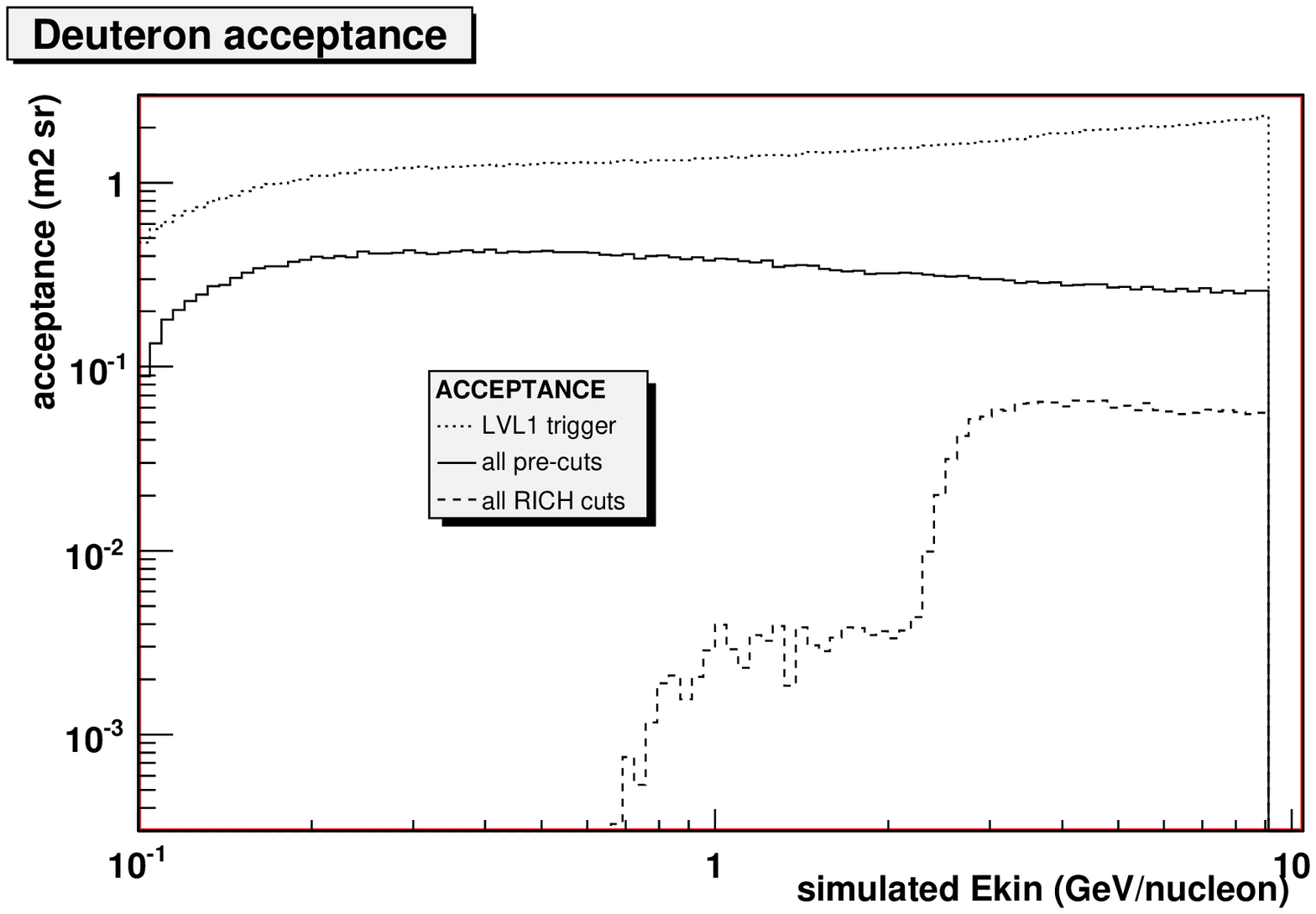,width=0.48\textwidth,clip=}}

\vspace{-0.3cm}

\caption{Acceptance for protons (top) and deuterons (bottom) at different
stages of event analysis. The lower line in each plot corresponds to the
final acceptance.
\label{pdaccep}}

\end{figure}

After all cuts, an acceptance of $\sim$~0.07 m${}^2\cdot$sr was
obtained for protons, and $\sim$~0.05 m${}^2\cdot$sr for deuterons at
\mbox{$E_{kin} >$~3 GeV/nucleon} (Fig.~\ref{pdaccep}). The increase by a factor
$\sim$~10 in the acceptance above the aerogel threshold reflects the
relative dimensions of the two radiators in the RICH detector.

The main background in the deuteron case comes from non-gaussian tails of
proton events with a bad velocity reconstruction. Errors in rigidity
reconstruction ($\Delta R / R \sim$~2\% in the GeV region) are not
critical for this case.

The specific set of cuts shown here corresponds to an example of a
selection procedure. Other variations are possible. In particular,
rejection factors may be improved by applying stricter cuts, at the
expense of a further acceptance reduction.

\section{Conclusions}

AMS-02 will provide a major improvement on the current knowledge of cosmic
rays. A total statistics of more than 10${}^{10}$ events will be collected
during its operation. Detailed simulations have been performed to evaluate
the detector's particle identification capabilities, in particular those
of the RICH. Simulation results show that the separation of light isotopes
is feasible. Using a set of simple cuts based on event data, relative mass
resolutions of $\sim$~2 \% and rejection factors higher than 10${}^4$ have
been attained in D/p separation at energies of a few GeV/nucleon. The
separation procedure presented here might be crucial for the identification
of an antideuteron flux resulting from neutralino annihilation.

\end{document}